\documentclass[12pt]{elsarticle}

\biboptions{authoryear,semicolon}

\usepackage{natbib}

\usepackage{graphicx}

\usepackage{amssymb}

\usepackage{lineno}

\begin{document}

\begin{frontmatter}






\title{Phase transitions in MgSiO$_3$ post-perovskite \\ in super-Earth mantles}


\author[affil1,affil2,affil3]{Koichiro Umemoto\corref{cor1}}
\ead{umemoto@elsi.jp}
\author[affil4,affil5,affil6]{Renata M. Wentzcovitch}
\author[affil3,affil7]{Shunqing Wu}
\author[affil3]{Min Ji}
\author[affil3]{Cai-Zhuang Wang}
\author[affil3]{Kai-Ming Ho}

\address[affil1]{Earth-Life Science Institute, Tokyo Institute of Technology,
Ookayama, Meguro-ku, Tokyo 152-8550, Japan}

\address[affil2]{Department of Earth Sciences, University of Minnesota, 310 Pillsbury drive SE, Minneapolis, MN 55455, USA}

\address[affil3]{Ames Laboratory, US DOE and Department of Physics and Astronomy, Iowa State University, Ames, Iowa 50011, USA}

\address[affil4]{Department of Applied Physics and Applied Mathematics, Columbia University, New York, NY, 10027}

\address[affil5]{Department of Earth and Environmental Sciences, Columbia University, New York, NY, 10027}

\address[affil6]{Lamont-Doherty Earth Observatory, Columbia University, Palisades, NY, 10964}

\address[affil7]{Department of Physics, Xiamen University, Xiamen 361005, China}


\cortext[cor1]{Corresponding author}
\fntext[fn1]{E-mail: umemoto@elsi.jp}








\begin{abstract}
The highest pressure form of the major Earth-forming mantle silicate is MgSiO$_3$ post-perovskite (PPv).
Understanding the fate of PPv at TPa pressures is the first step for understanding the mineralogy of super-Earths-type exoplanets, arguably the most interesting for their similarities with Earth.
Modeling their internal structure requires knowledge of stable mineral phases, their properties under compression, and major element abundances. Several studies of PPv under extreme pressures support the notion that a sequence of pressure induced dissociation transitions produce the elementary oxides SiO$_2$ and MgO as the ultimate aggregation form at $\sim$3 TPa.
 However, none of these studies have addressed the problem of mantle composition, particularly major element abundances usually expressed in terms of three main variables, the Mg/Si and Fe/Si ratios and the Mg\#, as in the Earth. Here we show that the critical compositional parameter,
 the Mg/Si ratio, whose value in the Earth's mantle is still debated, is a vital ingredient for modeling phase transitions and internal structure of
super-Earth mantles.
Specifically, we have identified new sequences of phase transformations, including new recombination reactions that depend decisively on this ratio.
 This is a new level of complexity that has not been previously addressed, but proves essential for modeling the nature and number of internal layers in these rocky mantles.
\end{abstract}


\end{frontmatter}

\newpage
\section{Introduction}

The incredible 1995 discovery of a Jupiter mass planet around a sun-like star, Pegasi 51
\citep{Mayor1995}, marked the dawn of a new age in planetary science and astronomy. Approximately 3,500 exoplanets have been confirmed since then, among which over 800 are super-Earths and nearly 400 are of terrestrial type, i.e., rocky planets. The discovery of these smaller rocky planets exploded with the deployment of the Kepler space telescope, which to date still has in its log $\approx$ 4,700 exoplanet candidates to be confirmed. Among these planets, the terrestrial ones are arguably the most interesting. Comparisons with Earth offer insights into our own planetary system formation process, and if found in the habitable zone, they might harbor life.
Kepler 62e and 62f\citep{Borucki2013},
GJ 667c\citep{Delfosse2013},
Kepler 186f\citep{Quintana2014},
Kepler 438b and 442b \citep{Torres2015},
Wolf 1061c \citep{Wright2016},
and three planets in the TRAPPIST-1 system \citep{Gillon2017} are remarkable examples of these potentially habitable planets.
Super-Earths and mini-Neptunes with masses up to $\sim$17 $M_\oplus$ are also remarkable because they seem to be relatively abundant, but none exist in our own solar system.  
These extraordinary discoveries are exciting public curiosity and interest and offer fertile grounds for frontier research.
Planetary modelers have been actively studying super-Earths for over a decade
\citep{Valencia2006,Fortney2007,Seager2007,Sotin2007,Grasset2009,Bond2010,Sasselov2010,Wagner2011,Dorn2015,Unterborn2016,Dorn2017,Unterborn2017}. 
The general description of these planets' compositions and internal structures have been attained with knowledge of their masses, radii, and the basic equations governing the physics of planets, among which, the equation of state of planet forming materials is essential. Modern approaches to materials theory and simulations also started being employed a decade ago to investigate the mineralogy of these planets\citep{Umemoto2006}. These calculations offer a level of detail that will be fundamental for understanding the dynamics of these planets. Mineral and aggregate phase diagrams, thermodynamics properties, and thermal conductivity are nowadays used as input for geodynamics simulations\citep{dave1,dave2} and will help answering questions regarding the nature of plate tectonics in these systems, a first order issue for habitability.
These {\it ab initio} studies of materials properties at ultra-high pressures expand the horizons of materials simulations and offer targets for novel high pressure experimental techniques reaching nearly TPa pressures - a new frontier in high pressure research. In this paper we address the fate of MgSiO$_3$, the major Earth forming phase, up to 4 TPa from a new perspective.
We focus on the important effect of the Mg/Si ratio 
of the lower mantle (Mg/Si$_{\rm LM}$)
as a key factor in determining phase equilibrium, therefore, the rocky mantle structure of super-Earths.

MgSiO$_3$ perovskite (bridgmanite) is the major constituent of the Earth's mantle and its highest pressure polymorph in the mantle is post-perovskite (PPv) \citep{Murakami2004,Oganov2004,Tsuchiya2004}. In super-Earths, other forms of aggregation of MgO and SiO$_2$ stabilized by the higher pressures and temperatures in their mantles are expected. Several {\it ab initio} computational studies have predicted dissociations of MgSiO$_3$ PPv as post-PPv transitions \citep{Umemoto2006,Umemoto2011,Wu2014,Niu2015}. These studies have shown that MgSiO$_3$ PPv should dissociate successively into $I\bar{4}2d$-type Mg$_2$SiO$_4$ + $P2_1/c$-type MgSi$_2$O$_5$, into Fe$_2$P-type SiO$_2$ + $I\bar{4}2d$-type Mg$_2$SiO$_4$, and then into the elementary oxides CsCl-type MgO + Fe$_2$P-type SiO$_2$
(see Supporting information for a brief explanation of the crystal structures).
 So far these post-PPv phase transitions have not been confirmed experimentally because of the extremely high pressures. In MgSiO$_3$ the Mg/Si ratio is 1, but the Earth's upper mantle - perhaps the whole mantle - has a Mg/Si ratio approximately equal to 1.25 and
stars with orbiting planets have stellar Mg/Si ratios varying very widely \citep{Adibekyan2015,Santos2015}.
Phase transitions and different aggregates may occur by varying Mg/Si$_{\rm LM}$ which impact the nature and number of internal layers in super-Earth mantles.
In the present paper, we study these phase transitions in super-Earths with different Mg/Si$_{\rm LM}$ ratios  
using {\it ab initio} calculations combined with the adaptive genetic algorithm (AGA) \citep{Ji2011,Wu2014} for structural search. We explore crystal structures with novel compositions in an unbiased manner without {\it a priori} assumptions of likely structural motifs. 

\section{Computational method}

The AGA method employs auxiliary model potentials to enable fast exploration of structural space. With this efficient method we were able to investigate supercells with 2$\sim$8 formula units for all compounds considered here (up to 88 atoms). For more details of the AGA method, see \citet{Ji2011,Wu2014}. 

For first-principles calculations, we adopted the local-density approximation \citep{PZ}.
To calculate total energies, forces and stresses during the AGA structural searches,
we used the following pseudopotentials. For Mg, a pseudopotential by von Barth-Car's method
\citep{Karki2000}. Five configurations, $3s^2 3p^0$, $3s^1 3p^1$, $3s^1 3p^{0.5} 3d^{0.5}$,
$3s^1 3p^{0.5}$, and $3s^1 3d^1$ with decreasing weights 1.5, 0.6, 0.3, 0.3,
and 0.2, respectively, were used.
Core radii for all quantum numbers $l$ are 2.5 a.u..
The pseudopotentials for Si and O were generated by Vanderbilt's method \citep{Vanderbilt1990}.
The valence electronic configurations used are
$3s^2 3p^1$ and $2s^2 2p^4$ for Si and O.
Core radii for all quantum numbers $l$ are 1.6 and 1.4 a.u. for Si and O.
The plane-wave cutoff energy is 40 Ry.
Brillouin-zone integration was performed
over {\bf k}-point meshes of spacing $2\pi\times 0.05$\AA$^{-1}$.
In order to refine candidate structures obtained by the AGA searches, we used a set of harder
pseudopotentials which are more suitable for ultrahigh pressures. 
They are the same as used in \citet{Umemoto2006,Umemoto2011}.
For these hard pseudopotentials, we used a cutoff energy of 400 Ry.
Structural relaxations were performed using variable-cell-shape molecular
dynamics \citep{Wentzcovitch1993}.
To test for structural
stability, phonon and vibrational density of states calculations
were carried out for candidate structures using density-functional-perturbation
theory \citep{Baroni2001}.
We computed the vibrational contribution to the free energies within the quasiharmonic approximation (QHA) \citep{Wallace1972}.
{\bf k}-point grid for electronic-structure calculations and {\bf q}-point grid for QHA
were ($6\times 6\times 2, 12\times 12\times 10$) for MgSiO$_3$ PPv,
($4\times 4\times 4, 12\times 12\times 12$) for NaCl-type MgO,
($8\times 8\times 8, 16\times 16\times 16$) for CsCl-type MgO,
($4\times 4\times 6, 16\times 16\times 16$) for Fe$_2$P-type SiO$_2$, and
($4\times 4\times 4, 12\times 12\times 12$) for $I\bar{4}2d$-type Mg$_2$SiO$_4$, and
($4\times 2\times 2, 10\times 8\times 10$) for $P2_1/c$-type MgSi$_2$O$_5$.
All first-principles calculations were performed using the Quantum-ESPRESSO
\citep{Giannozzi2009}, which has been interfaced with the GA scheme
in a fully parallel manner.

\section{Results and Discussion}

\subsection{Successive dissociations of MgSiO$_3$}
\begin{figure}[h]
\centering
\includegraphics[width=.9\linewidth]{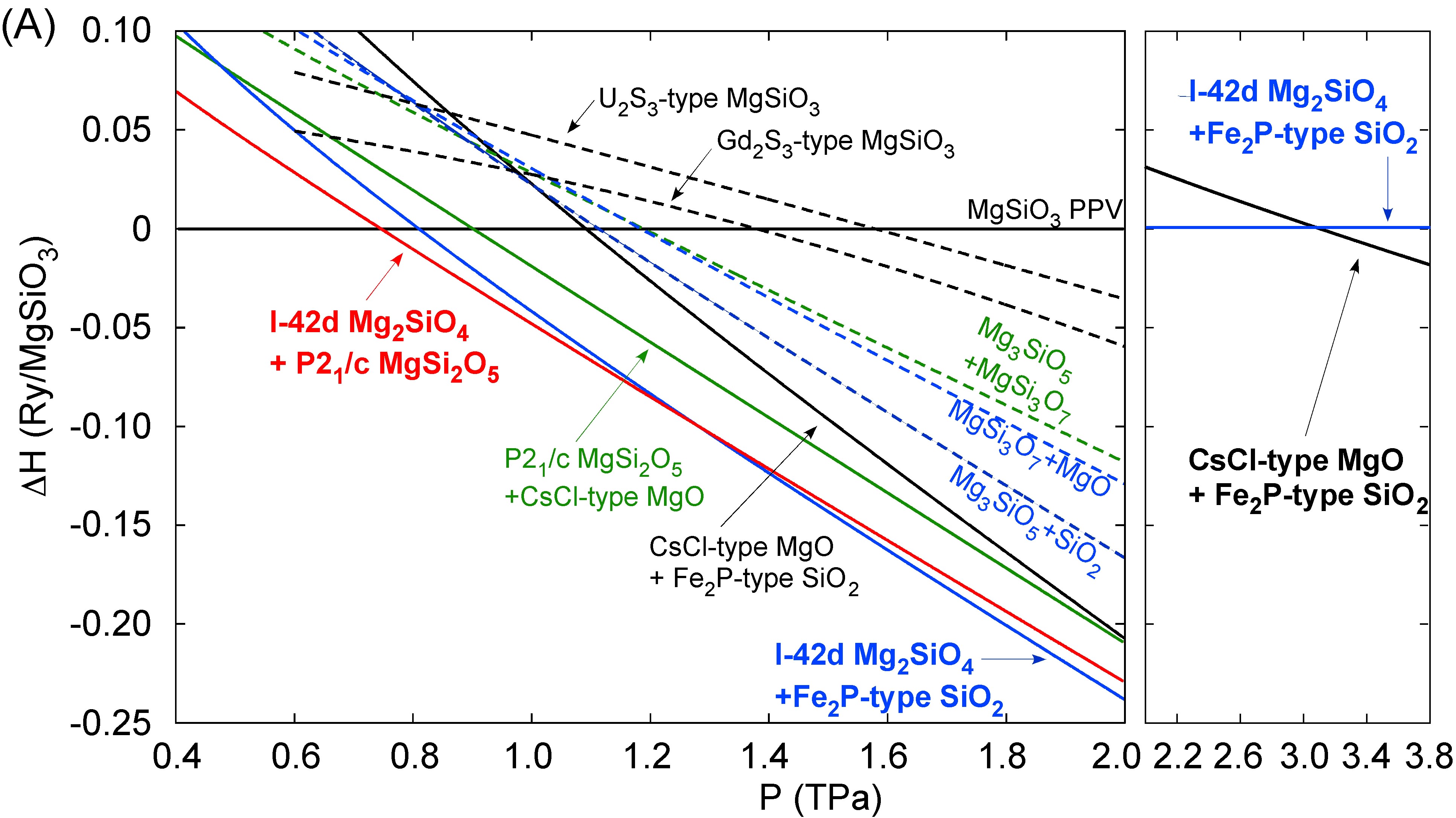}
\caption{Relative enthalpies of aggregation of possible dissociation products of MgSiO$_3$ and crystalline post-PPv MgSiO$_3$ with respect to MgSiO$_3$ PPv.}
\label{dH}
\end{figure}

\begin{figure}[h]
\centering
\includegraphics[width=.5\linewidth]{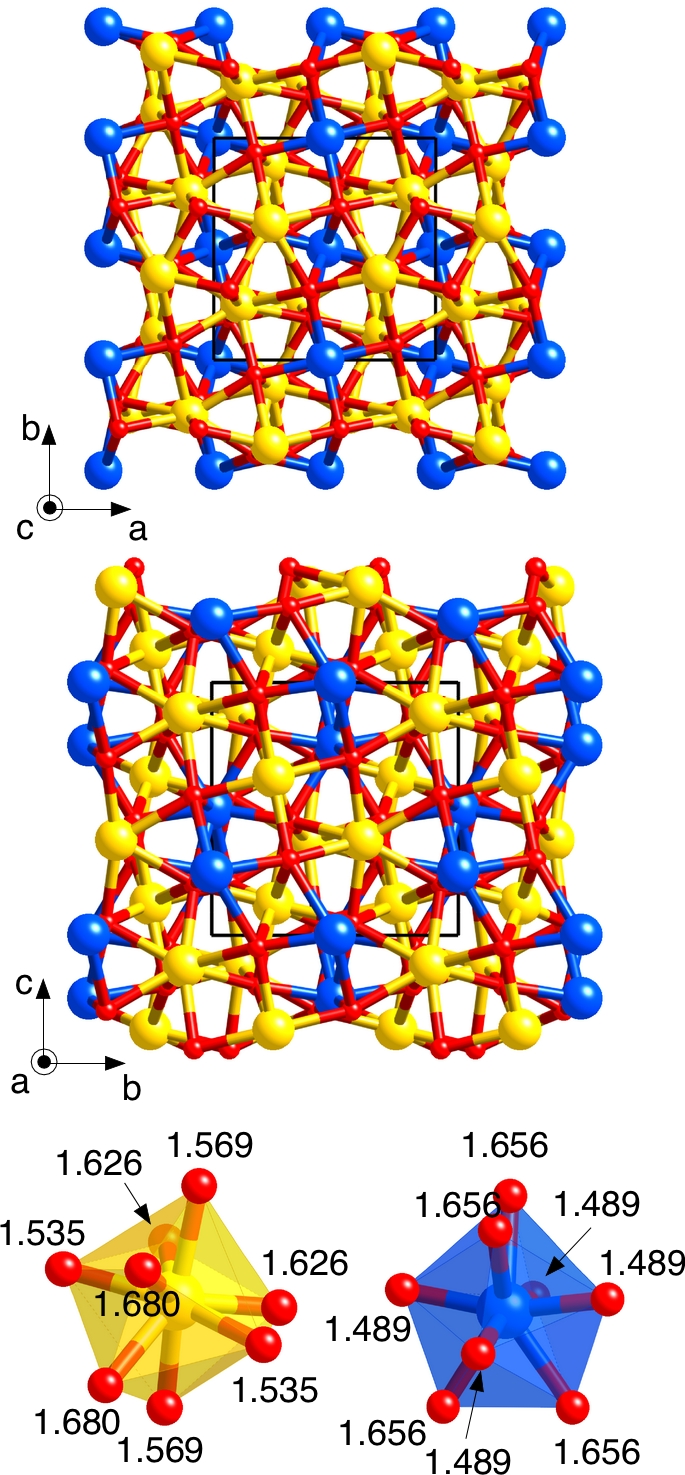}
\caption{Crystal structure of $I\bar{4}2d$-type Mg$_2$SiO$_4$. Yellow, blue, and red spheres denote Mg, Si and O atoms.
Lattice constants at 1 TPa are $(a,c)$=(4.555\AA, 4.671\AA).
Atomic coordinates are Mg $(8d) (0.11348, 1/4, 1/8)$, Si $(4b) (0, 0, 1/2)$, and O $(16e) (0.93007, 0.81560, 0.79792)$.
Bond lengths in Mg and Si polyhedra at 1 TPa are given in units of \AA.}
\label{structure}
\end{figure}

As shown in previous studies \citep{Wu2014,Niu2015}, MgSiO$_3$ PPv undergoes successive dissociations: MgSiO$_3$ PPv $\to$ $I\bar{4}2d$-type Mg$_2$SiO$_4$ + $P2_1/c$-type MgSi$_2$O$_5$ at 0.75 TPa $\to$ $I\bar{4}2d$-type Mg$_2$SiO$_4$ + Fe$_2$P-type SiO$_2$ (i.e., dissociation of MgSi$_2$O$_5$) at 1.31 TPa $\to$ CsCl-type MgO + Fe$_2$P-type SiO$_2$ at 3.10 TPa (see Fig.~\ref{dH}). It is worthwhile commenting on the crystal structure of $I\bar{4}2d$-type Mg$_2$SiO$_4$ (Fig.~\ref{structure}). For similar comments on $P2_1/c$-type MgSi$_2$O$_5$, see \citet{Umemoto2011}. $I\bar{4}2d$-type Mg$_2$SiO$_4$ is a body-centered-tetragonal phase. As far as we know, this structure has not been identified in any other substance. However, its cation configuration is identical to that of Zn$_2$SiO$_4$-II whose space group is $I\bar{4}2d$ also \citep{Marumo1971}. The difference between $I\bar{4}2d$-type Mg$_2$SiO$_4$ and Zn$_2$SiO$_4$-II structures is in the oxygen sub-lattice. Both Mg and Si atoms are eight-fold coordinated, while Zn and Si atoms in Zn$_2$SiO$_4$-II are four-fold tetrahedrally coordinated. The structure of $I\bar{4}2d$-type Mg$_2$SiO$_4$ might be a viable high-pressure form of Zn$_2$SiO$_4$. Lattice dynamics calculations indicate that $I\bar{4}2d$-type Mg$_2$SiO$_4$ is dynamically stable at least up to 4 TPa. While Mg polyhedra share their faces, Si polyhedra share their edges. The eight-fold coordinated Mg and Si polyhedra are somewhat intermediate between NaCl-type octahedra and CsCl-type cubes. This type of cation site contrasts in $P2_1/c$-type MgSi$_2$O$_5$ where Mg and half of Si polyhedra form tricapped triangular prisms, which are structural units of Fe$_2$P-type SiO$_2$ \citep{Tsuchiya2011,Wu2011}. The other half of Si sites are octahedrally coordinated with one extra oxygen attached. This is similar to the case of pyrite-type SiO$_2$. Thermodynamic quantities of $I\bar{4}2d$-type Mg$_2$SiO$_4$ obtained using the quasiharmonic approximation (QHA) are shown in Table \ref{quantities}; similar quantities for $P2_1/c$-type MgSi$_2$O$_5$ were published in a previous study \citep{Umemoto2011}.

\begin{figure}[h]
\centering
\includegraphics[width=.9\linewidth]{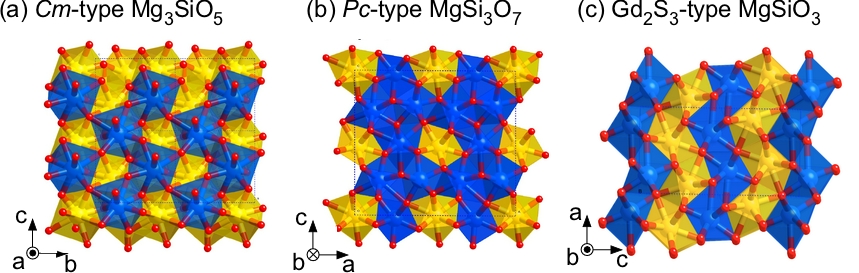}
\caption{Crystal structures of (a) $Cm$-type Mg$_3$SiO$_5$, 
(b) $Pc$-type MgSi$_3$O$_7$ and
(c) Gd$_2$S$_3$-type MgSiO$_3$. Structural parameters are given in Table \ref{struct_param2}. }
\label{Struct2}
\end{figure}

This dissociation of MgSiO$_3$ into Mg$_2$SiO$_4$, a MgO-rich compound, and MgSi$_2$O$_5$, a SiO$_2$-rich one, suggests the possibility of further dissociations into other MgO- and SiO$_2$-``richer'' compounds, i.e., (MgO)$_{m}$(SiO$_2$) with $m_1\ge3$ and (MgO)(SiO$_2$)$_{n}$ with $n_1\ge3$. To check this possibility, we performed structural searches for Mg$_3$SiO$_5$ ($m=3$) and MgSi$_3$O$_7$ ($n=3$) with the AGA. The lowest-enthalpy structures of Mg$_3$SiO$_5$ and MgSi$_3$O$_7$ are monoclinic with $Cm$ and $Pc$ symmetries, respectively (see Fig.~\ref{Struct2}).
As shown in Fig.~\ref{dH}, this aggregate of Mg$_3$SiO$_5$ and MgSi$_3$O$_7$ is metastable with respect to that of Mg$_2$SiO$_4$ and MgSi$_2$O$_5$. The possibility of other higher pressure forms of MgSiO$_3$ was also investigate with the AGA. The lowest-enthalpy crystalline post-PPv phase of MgSiO$_3$ has Gd$_2$S$_3$-type structure (Fig.~\ref{Struct2}). The Gd$_2$S$_3$-type phase is a typical high-pressure structure in $A_2$O$_3$ type compounds({\it e.g.} \citet{Umemoto2010} and references therein). Gd$_2$S$_3$-type MgSiO$_3$ has lower enthalpy than U$_2$S$_3$-type, which was predicted to be the stable form of Al$_2$O$_3$ at ultra-high pressures \citep{Umemoto2008}. Gd$_2$S$_3$-type MgSiO$_3$ is also metastable compared to the aggregate of Mg$_2$SiO$_4$ and MgSi$_2$O$_5$ (Fig.~\ref{dH}). Therefore, we exclude the possibilities of reactions involving (MgO)$_{m}$(SiO$_2$) with $m\ge3$, (MgO)(SiO$_2$)$_{n}$ with $n\ge3$, and crystalline post-PPv MgSiO$_3$. We conclude that the vital compounds relevant to interiors of super-Earths are MgO, SiO$_2$, MgSi$_2$O$_5$, and Mg$_2$SiO$_4$.

\subsection{Recombination of MgSiO$_3$ with MgO or SiO$_2$}

\begin{figure}[h]
\centering
\includegraphics[width=.9\linewidth]{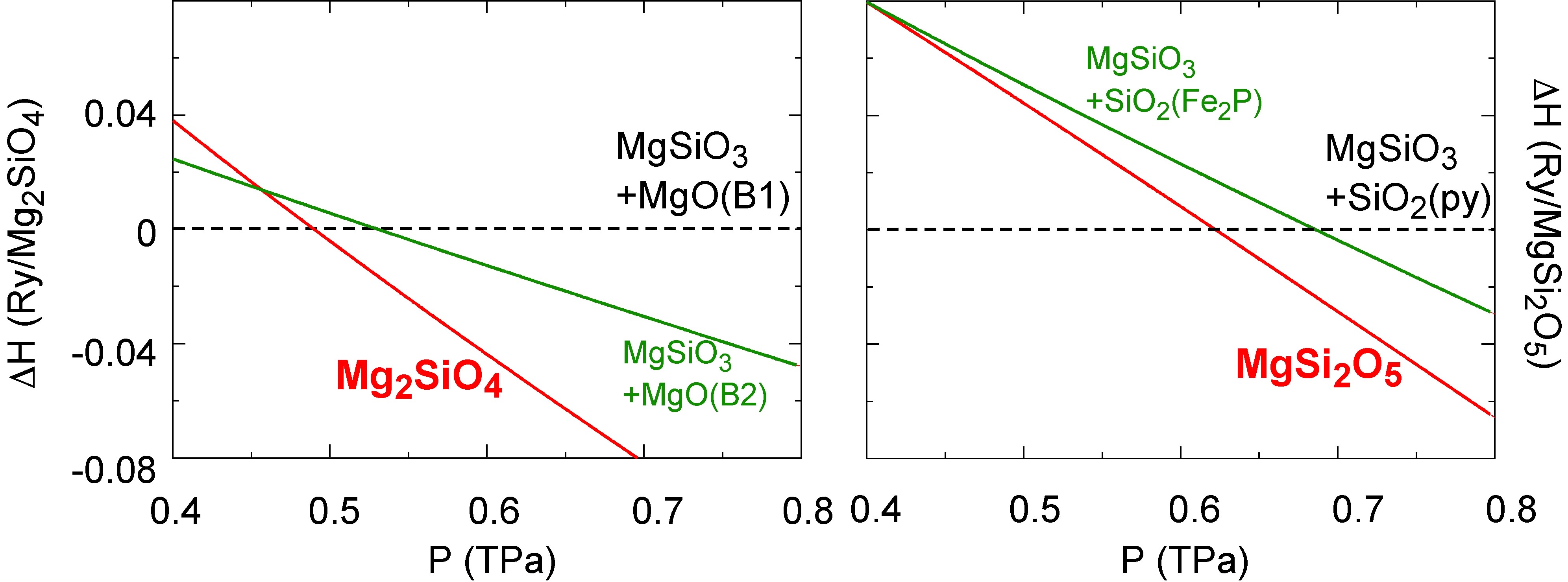}
\caption{Relative enthalpies of $I\bar{4}2d$-type Mg$_2$SiO$_4$ with respect to aggregations of MgSiO$_3$ and MgO (left) and $P2_1/c$-type MgSi$_2$O$_5$ with
respect to aggregations of MgSiO$_3$ and SiO$_2$ (right).}
\label{dH_recombination}
\end{figure}

Now we discuss the cases where MgSiO$_3$ PPv coexists with MgO or with SiO$_2$, which is expected in mantles with Mg/Si$_{\rm LM}$ ratio 
$>$ 1 and $<$ 1, respectively. The low-pressure form of MgO, NaCl-type, transforms to the CsCl-type phase \citep{Coppari2013}. The static transition pressure of 0.53 TPa is consistent among {\it ab initio} calculations ({\it e.g.} \citet{Mehl1988}).
QHA free energy calculations indicate that NaCl-type MgO and MgSiO$_3$ PPv combine to form $I\bar{4}2d$-type Mg$_2$SiO$_4$ beyond 0.49 TPa (see Fig.~\ref{dH_recombination}(A)). This is surprising because in the Earth Mg$_2$SiO$_4$ ringwoodite, a major transition zone phase, dissociates into MgO periclase and MgSiO$_3$ bridgmanite. In the Earth's mantle these phases contain small amounts of Fe, but the dissociation also exists in the Fe-free compound. This dissociation transition is responsible for the major ``660 km'' velocity discontinuity defining the upper boundary of the lower mantle. Here we see that under much higher pressures corresponding to the deep interiors of super-Earths, NaCl-type MgO and MgSiO$_3$ PPv recombine into $I\bar{4}2d$-type Mg$_2$SiO$_4$. Similarly, MgSiO$_3$ PPv and pyrite-type SiO$_2$ combined into $P2_1/c$-type MgSi$_2$O$_5$ at 0.62 TPa (see Fig.~\ref{dH_recombination}(B)). These recombination pressures are lower than the dissociation pressure of MgSiO$_3$ PPv into Mg$_2$SiO$_4$ and MgSi$_2$O$_5$ and also lower than the transition pressures of NaCl-type to CsCl-type MgO and of pyrite-type to Fe$_2$P-type SiO$_2$.
A new view of super-Earth mantles then emerges.

\section{Dissociation and recombination in super-Earths}

\begin{figure}[h]
\centering
\includegraphics[width=.9\linewidth]{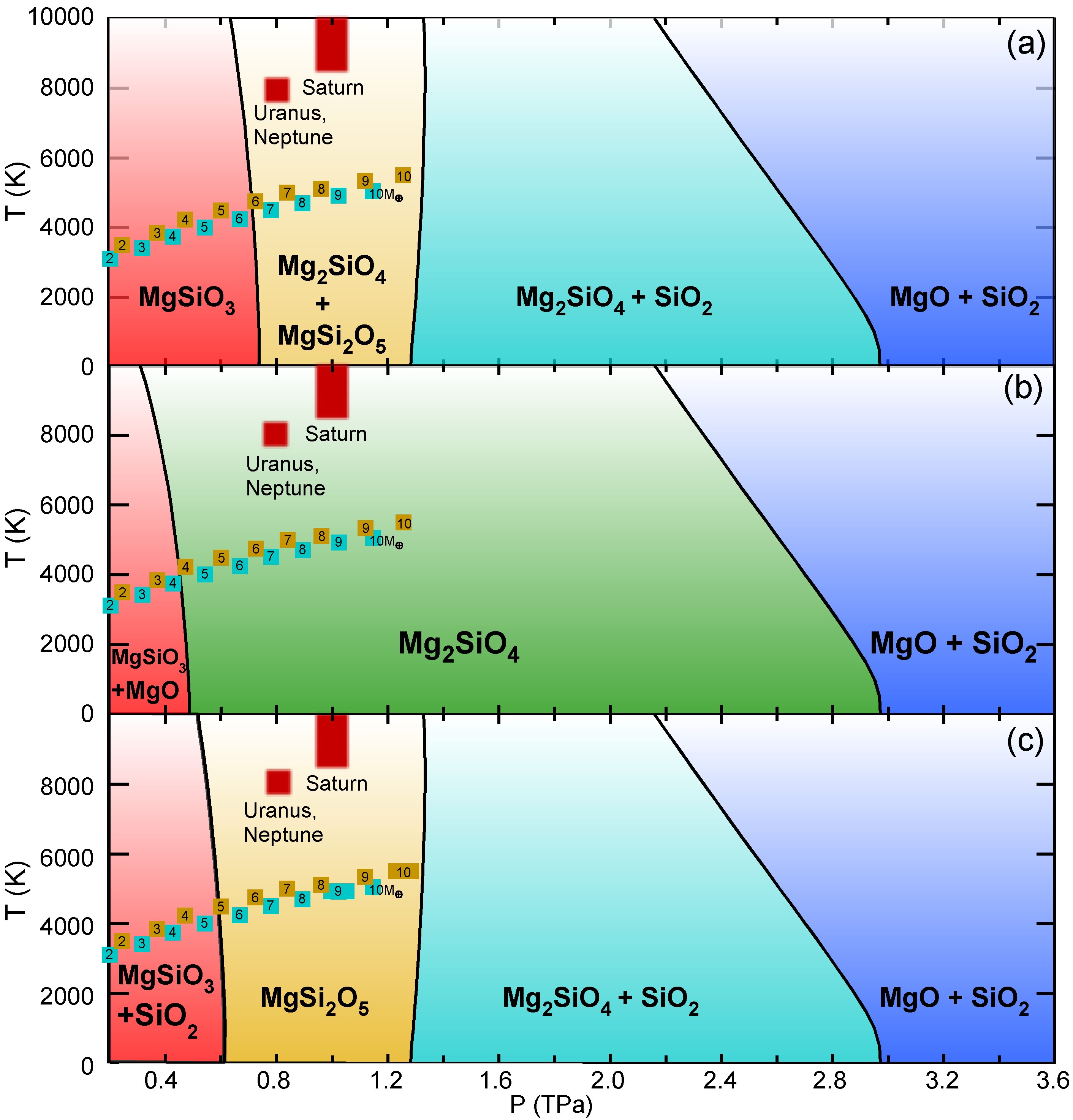}
\caption{Phase diagram showing (a) three-stage of dissociations of MgSiO$_3$ PPv, (b) recombination of MgSiO$_3$ PPv and NaCl-type MgO, and (c) recombination of MgSiO$_3$ and pyrite-type SiO$_2$;
in (b) and (c), the molar ratios between MgSiO$_3$ and MgO/SiO$_2$ are 1:1.
Red spots denote estimated pressure-temperature conditions at core-envelope boundaries
in the solar giants (Jupiter's condition, $\sim$4 TPa and $\sim$15,000--20,000 K,
is not shown here) \citep{Guillot2004}.
Brown and light blue squares represent pressure-temperature conditions at CMB in terrestrial and ocean exoplanets, assuming their Earth-like chemical compositions${}^\dagger$ \citep{Sotin2007}. Numbers in squares indicate planet masses in units of Earth mass (M$_{\oplus}$).
}
\label{PB}
\end{figure}

The resulting sequence of phase transitions can be examined by computing aggregate phase diagrams using the QHA.
For pure MgSiO$_3$, the sequence of dissociations predicted by static calculations \citep{Wu2014} holds up to 10,000 K (see Fig.~\ref{PB}a). This result contrasts with a recent prediction \citep{Niu2015} showing that a different sequence of dissociation reactions, namely Mg$_2$SiO$_4$ + MgSi$_2$O$_5$ $\to$ 
3/2 (MgSi$_2$O$_5$ + MgO)
occurs above $\sim$ 6,000 K. For super-Earths and ocean exoplanets with masses of 1--10 $M_\oplus$, CMB pressures and temperatures have been estimated
assuming Earth-like mantle compositions and more basic thermal equations of state \citep{Valencia2006,Sotin2007}${}^\dagger$.
For mantles with Mg/Si ratio equal to 1, MgSiO$_3$ PPv survives in terrestrial and ocean super-Earths with masses smaller than $\sim 6M_\oplus$. In the presence of MgO the recombination of MgSiO$_3$ and MgO occurs when their masses are larger than $\sim 4M_\oplus$ (see Fig.~\ref{PB}b) while in the presence of SiO$_2$ the recombination transition is expected in planets with masses larger than $\sim 5M_\oplus$ (see Fig.~\ref{PB}c).
The first dissociation into Mg$_2$SiO$_4$ and MgSi$_2$O$_5$ should occur in super-Earths with masses larger than $\sim 6M_\oplus$, for example Kepler-20b \citep{Gautier2012} and possibly CoRoT-7b whose mass has been under debate
\citep{Leger2009,Hatzes2011}. 
The Clapeyron slopes
(slope of the phase boundary which is given by $dP/dT$ on the phase boundary)
of three successive dissociations in MgSiO$_3$ at 5,000 K (see Fig.~\ref{PB}a) are $-10$, $+6$, and -92 MPa/K, respectively.
Those of the two recombination transitions, i.e., MgSiO$_3$ + MgO $\to$ Mg$_2$SiO$_4$ and MgSiO$_3$ + SiO$_2$ $\to$ MgSi$_2$O$_5$ are $-16$ and $-9$ MPa/K, respectively.
Except for the dissociation of MgSi$_2$O$_5$ $\to$ Mg$_2$SiO$_4$ + SiO$_2$ (Fig.~\ref{PB}a and c), the Clapeyron slopes of all phase boundaries are negative. Recombination and dissociation with large negative Clapeyron slopes in the middle of a silicate mantle may promote compositional layering \citep{Christensen1985,Peltier1992,Tackley1995}. In Mars, the post-spinel transition with negative Clapeyron slope near the CMB is believed to be the cause of a proposed large superplume \citep{Weinstein1995}. Similarly, in super-Earths with masses $\sim 4M_\oplus$ (or $\sim 6M_\oplus$), recombination of NaCl-type MgO and MgSiO$_3$ (or dissociation of MgSiO$_3$) could induce a comparable phenomenon.

\begin{figure}[h]
\centering
\includegraphics[width=.8\linewidth]{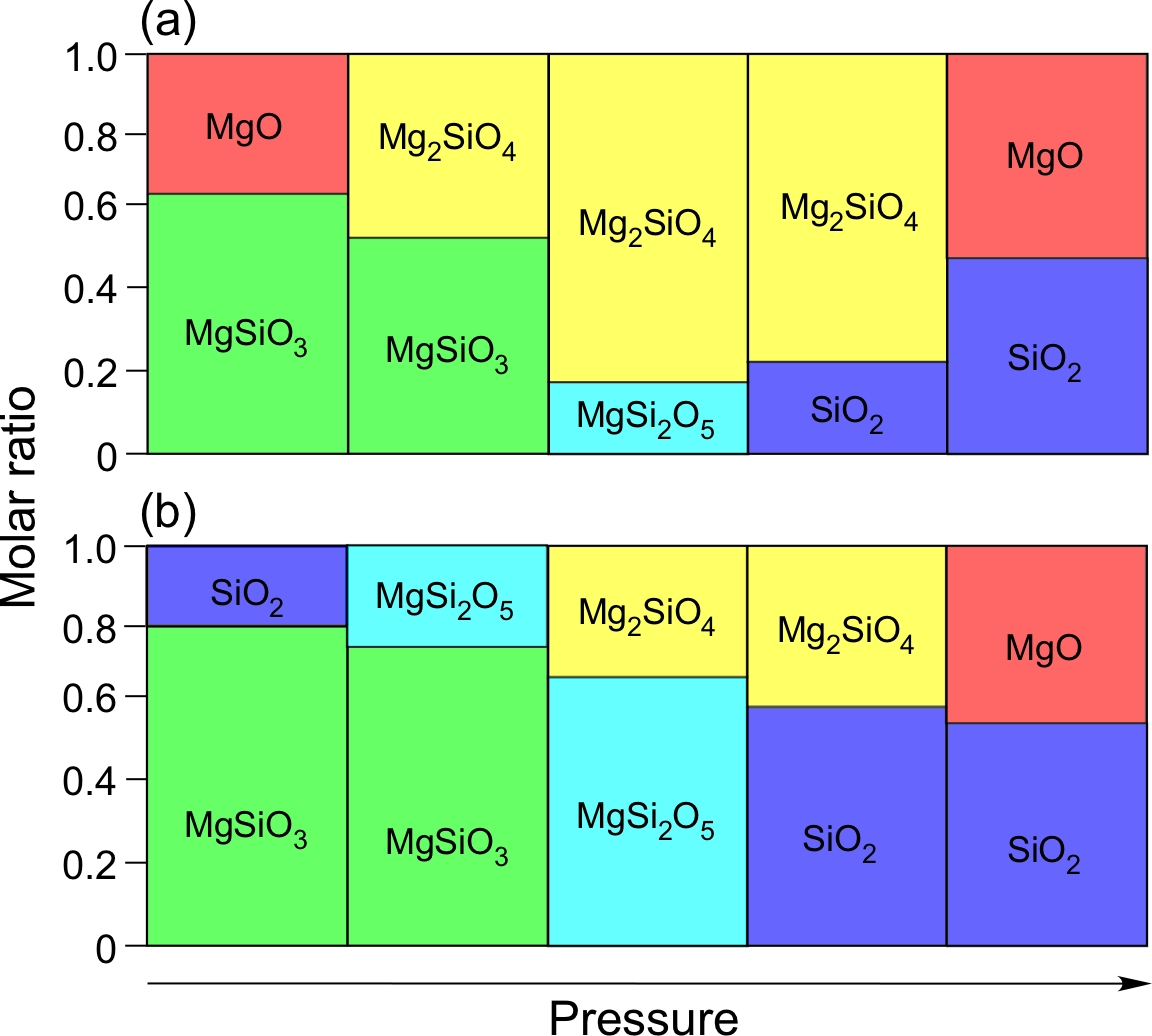}
\caption{Pressure-induced changes of molar ratio of components of super-Earth's mantle under pressure. (a) In a case of coexistence of MgSiO$_3$ and MgO. The molar ratio of MgSiO$_3$ and MgO is assumed to be 0.65:0.35, corresponding to Mg/Si$_{\rm LM}$=1.54. (b) In a case of coexistence of MgSiO$_3$ and SiO$_2$. The molar ratio of MgSiO$_3$ and SiO$_2$ is assumed to be 0.80:0.20, corresponding to Mg/Si$_{\rm LM}$=0.8.}
\label{molarratio}
\end{figure}

Stellar Mg/Si ratios can vary substantially around 1 \citep{Adibekyan2015,Santos2015} and the number of transitions up to 3 TPa increases in both Mg-rich and Si-rich mantles, in which MgSiO$_3$ coexists with MgO and SiO$_2$ respectively, because of the newly found recombination reactions.
For Mg-rich mantles, by assuming the molar ratio between MgSiO$_3$ PPv and MgO to be $x:y$(=1/(Mg/Si$_{\rm LM}-$1)), the sequence of transitions is as follows (Fig.~\ref{molarratio}(a)):
\begin{eqnarray*}
x{\rm MgSiO}_3 + y {\rm MgO} &\to& (x-y){\rm MgSiO}_3 + y {\rm Mg}_2{\rm SiO}_4 \\
 &\to& \frac{x+2y}{3} {\rm Mg}_2{\rm SiO}_4 +\frac{x-y}{3} {\rm MgSi}_2{\rm O}_5 \\
 &\to& \frac{x+y}{2}{\rm Mg}_2{\rm SiO}_4 +\frac{x-y}{2}{\rm SiO}_2 \\
 &\to& (x+y){\rm MgO}+x{\rm SiO}_2.
\end{eqnarray*}

It should be noted that NaCl-type MgO disappears with the recombination transition and MgO emerges again in the CsCl-type form with the last dissociation transition. Earth-like mantle compositions possibly vary from pyrolitic \citep{McDonough1995,Jackson1998,daSilva2000,Karki2001,WentzcovitchPRL2004,Irifune2010,Wang2015} to near-chondritic \citep{Murakami2012}. In a 
pyrolitic mantle 
$x:y \approx 0.65:0.35$ (if Fe is replaced by Mg, Ca by Mg, and Al by a (Mg,Si) coupled substitution). In this case, 67\% MgSiO$_3$ PPv combines with NaCl-type MgO into Mg$_2$SiO$_4$, while 33\% MgSiO$_3$ PPv survives and MgO disappears; therefore, the transition between NaCl- and CsCl-type MgO should not occur in super-Earths. The resulting molar fractions of MgSiO$_3$ and Mg$_2$SiO$_4$ are 0.46 and 0.54, respectively. After the first dissociation the molar fractions of Mg$_2$SiO$_4$ and MgSi$_2$O$_5$ are 0.82 and 0.18. For a near-chondritic mantle, the molar fractions of MgSiO$_3$ and MgO are $\sim$ 0.95 and 0.05. After the recombination the molar fractions of Mg$_2$SiO$_4$ and MgSiO$_3$ are 0.053 and 0.947, respectively, and after the first dissociation, the molar fractions of Mg$_2$SiO$_4$ and MgSi$_2$O$_5$ are 0.538 and 0.462.
For Si-rich mantles, where MgSiO$_3$ PPv coexists with pyrite-type SiO$_2$. By assuming the molar ratio between MgSiO$_3$ PPv and SiO$_2$ to be $x:y(=$1/(Si/Mg$_{\rm LM}-$1)), the transitions are (Fig.~\ref{molarratio}(b)):
\begin{eqnarray*}
x{\rm MgSiO}_3 + y {\rm SiO}_2 &\to& (x-y){\rm MgSiO}_3 + y {\rm MgSi}_2{\rm O}_5 \\
 &\to& \frac{x-y}{3} {\rm Mg}_2{\rm SiO}_4 +\frac{x+2y}{3} {\rm MgSi}_2{\rm O}_5 \\
 &\to& \frac{x}{2}{\rm Mg}_2{\rm SiO}_4 +\frac{x+2y}{2}{\rm SiO}_2 \\
 &\to& x{\rm MgO}+(x+y){\rm SiO}_2.
\end{eqnarray*}
Pyrite-type SiO$_2$ disappears after the recombination and SiO$_2$ emerges again in the Fe$_2$P-type phase after the second dissociation. In overall, these recombination reactions produce one extra mantle layer (see Fig.~\ref{PB}) whose contrast in properties depend on how much Mg/Si$_{\rm LM}$
departs from 1.

\section{Summary}

This {\it ab initio} study addressing the stability of MgSiO$_3$ PPv in super-Earth mantles with Mg/Si ratios
of the lower mantle
different from 1 identified new recombination reactions in both Mg- and Si-rich mantles in addition to the previously reported dissociation reactions. These recombinations with negative Clapeyron slopes introduce an extra layer in the mantles of super-Earths with masses larger than $\sim 4M_\oplus$ if Mg-rich or $\sim 5M_\oplus$ if Si-rich. Structural searches using the adaptive genetic algorithm revealed that dissociations of MgSiO$_3$ into products such as (MgO)$_m$(SiO$_2$) and (MgO)(SiO$_2$)$_n$ should not be expected and the relevant phases are those previously identified. Because of the recombination reaction, the NaCl- to CsCl-type transition in MgO does not occur
in super-Earth mantles.
CsCl-type MgO appears only with the final break down of Mg$_2$SiO$_4$ into pure oxides at much higher pressures and temperatures unlikely to materialize in mantles of super-Earths and mini-Neptunes. The last solid-solid transition identified so far remains the dissociation of Mg$_2$SiO$_4$ into the pure oxides Fe$_2$P-type SiO$_2$ and CsCl-type MgO at 3 TPa at low temperatures. This state of aggregation is the relevant one for the core of Jupiter like planets. Phase boundaries and thermodynamic quantities reported here should be important for modeling super-Earth interiors. 
Thermal conductivity is another important quantity for modeling super-Earth interiors, related to possibility of geodynamo and therefore implications for the habitability. First principles calculation of thermal conductivity should be performed, based on phase transitions we studied.

Although mantles of rocky planets more massive than Earth are expected to have achieved larger temperatures during formation, have ex-solved more Fe, and therefore are expected to have larger Mg\# (=[Mg]/[Mg+Fe]) than the Earth's mantle, searches for silicate and oxide mineral phases with Fe$_n$O$_m$ components should be undertaken in the future for a better understanding of these planets.

\section*{Footnote}
${}^\dagger$Mantle compositions and the size of core can be in first order specified by the relative abundances of Mg, Si, and Fe. \citet{Sotin2007} estimated pressure and temperature profiles in super-Earths by assuming Earth-like chemical compositions and equations of states of silicates in the mantle and iron alloys in the core;
Mg/Si=1.131, Fe/Si=0.986, and Mg\#=0.9.
Here we invoke CMB pressures and temperatures in super-Earths as reported in \citet{Sotin2007} as reference for the discussion on the effect of dissociation and recombination reactions in the mantle of these planets.
However, it should be noted here that, if super-Earths have chemical compositions rather different from the Earth, their CMB pressures and temperatures should be different from those estimated in \citet{Sotin2007}.
A mass-radius relation study for 65 exoplanets showed that average density of exoplanets increases when their radii ($R_P$) are smaller than 1.5 Earth's radius ($R_\oplus$) but decreases when $1.5R_\oplus <R_P<4R_\oplus$ \citep{Weiss2014}; it suggests that chemical compositions of super-Earths with different radii may be different from those of the Earth.

\begin{flushleft}
{\bf Acknowledgments}
\end{flushleft}
KU and RMW were supported by grants NSF/EAR 1047629, 1319368, and 1348066. KU is the main contributor to phonon and quaisharmonic computations using facilities at the Minnesota Supercomputing Institute and at the Laboratory for Computational Science and Engineering at the University of Minnesota and the ELSI supercomputing system at Tokyo Institute of Technology. SQW is the main contributor to the GA structure search performed at Ames Laboratory and SQW also acknowledges financial support from the National Natural Science Foundation of China (No. 11004165).
Work at Ames Laboratory (KU, SQW, JM, CZW and KMH) was supported by the US Department of Energy, Office of Science, Basic Energy Sciences, Materials Science and Engineering Division, including a grant of computer time at the National Energy Research Scientific Computing Centre (NERSC) in Berkeley, CA. Ames Laboratory is operated for the U.S. DOE by Iowa State University under contract \#DE-AC02-07CH11358.

\begin{flushleft}
{\bf References}
\end{flushleft}

\clearpage

\clearpage
\begin{center}
{\bf \large Supporting Information}
\end{center}

\normalsize
\begin{flushleft}
{\bf Crystal structures of MgO, SiO$_2$, and MgSiO$_3$ at ultrahigh pressures}
\end{flushleft}

In this paper, knowledge of some crystal structures is necessary.
Crystal structures of Mg$_2$SiO$_4$ and MgSi$_2$O$_5$ is given already in the main text.
Below we give a brief summary of other structures relevant to this study.
Figure S1 shows their crystal structures.
\begin{flushleft}
{\bf 1. MgO}
\end{flushleft}
At ambient condition, the crystal structure of MgO is NaCl-type, i.e., rock-salt type. Mg and O sublattices are face-centered cubic. They interpenetrate each other. Mg atoms are at the center of regular octahedra by O atoms and six-fold coordinates (Fig.~S1(a)).
At ultrahigh pressures, it transforms to a CsCl-type structure.
In this phase, Mg and O sublattices are simple cubic. 
They interpenetrate each other. Mg atoms are at the center of regular cube by O atoms and eight-fold coordinates (Fig.~S1(b)).

\begin{flushleft}
{\bf 2. SiO$_2$}
\end{flushleft}

SiO$_2$ is experimentally known to undergo a very complex pressure-induced phase transitions:
$\alpha$-quartz$\to$coesite$\to$CaCl$_2$-type$\to \alpha$-PbO$_2 \to$pyrite-type.
In $\alpha$-quartz and coesite, the structural unit is a SiO$_4$ tetrahedron.
In the phases between stishovite and pyrite-type, it is a SiO$_6$ octahedron.
The pyrite phase (Fig.~S1(c)) is the highest pressure form of SiO$_2$ among phases experimentally identified. 
Theoretically, Fe$_2$P-type structure has been predicted to be beyond pyrite.
In this structure, the structural unit is SiO$_9$; the coordination number of silicon is 9 (Fig.~S1(d)).
The SiO$_9$ polyhedron consists of a triangular prism of six O atoms and extra three O atoms are placed at some distance from three sides of the prism.
This kind of structural unit is frequently found in other AX$_2$-type compounds.

\begin{flushleft}
{\bf 3. MgSiO$_3$ post-post-perovksite}
\end{flushleft}

It is well known that post-perovsktie (PPv) is the final form of MgSiO$_3$
in the lower mantle of the Earth. As phase transitions beyond PPv, we predict
recombinations and dissociations in the present study.
Then, what if recombination and/or dissociation does not occur? In such a case,
MgSiO$_3$ PPv might transform to another crystalline phase.
There are two candidates of crystalline MgSiO$_3$ post-PPv: Gd$_2$S$_3$-type and U$_2$S$_3$-type.
Crystal structures of PPv, Gd$_2$S$_3$-type and U$_2$S$_3$-type are rather related to each other.
PPv consists of SiO$_6$-octahedron layers intercalated by Mg atoms. In Gd$_2$S$_3$- and U$_2$S$_3$-type phases, one O atom is newly attached to the SiO$_6$ octahedron. As a result, Si atoms are seven-fold coordinated at middle of SiO$_7$ polyhedra.
The Gd$_2$S$_3$-type phase consists of SiO$_7$ layers intercalated by Mg atoms (Fig.~S1(e)).
In the U$_2$S$_3$-type phase, SiO$_7$ layers are slightly buckled and connected to eath other. Mg atoms exist at voids of the SiO$_7$ network (Fig.~S1(f)).

\renewcommand{\thefigure}{S\arabic{figure}}
\setcounter{figure}{0}

\begin{figure}[h]
\centering
\includegraphics[width=.9\linewidth]{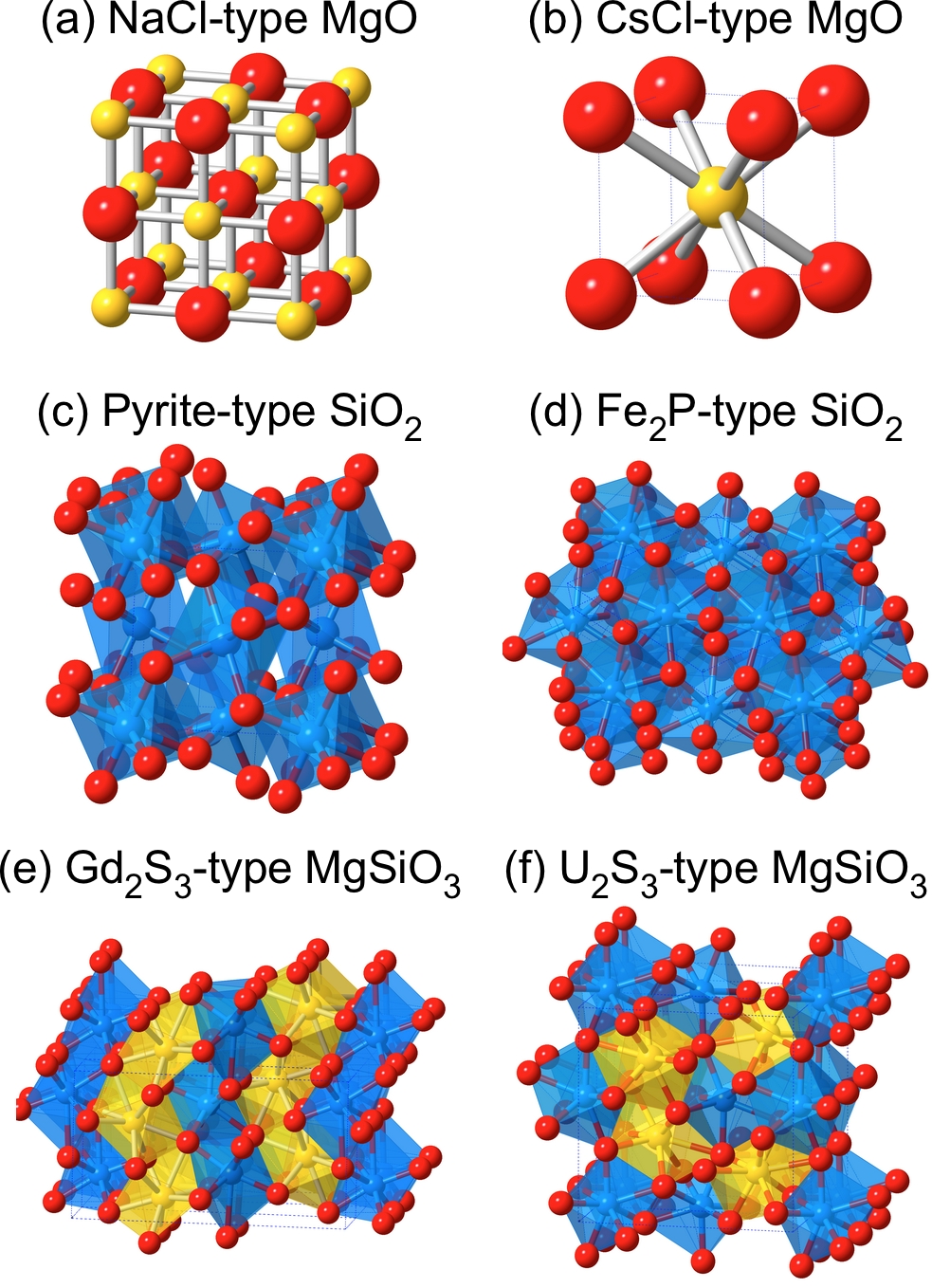}
\caption{Crystal structures of MgO, SiO2, and MgSiO3 relevant for this study.}
\label{SFig}
\end{figure}

\clearpage

\renewcommand{\thetable}{S\arabic{table}}
\setcounter{table}{0}

\begin{table*}[h]
\tiny
\caption{ Thermodynamic properties of $I\bar{4}2d$-type Mg$_2$SiO$_4$: density ($\rho$), thermal expansivity ($\alpha$), isothermal bulk modulus ($K_T$), adiabatic bulk modulus ($K_S$), constant-volume heat capacity ($C_V$), constant-pressure heat capacity ($C_P$), and Gr\"uneisen parameter ($\gamma$).
}
\begin{tabular}{cccccccc}
\hline
\multicolumn{8}{l}{1000 K} \\
$P$ (TPa) & $\rho$ (g/cm$^3$) & $\alpha$ ($10^{-5}$/K) & 
$K_T$ (TPa) & $K_S$ (TPa) & $C_V$ (J/mol/K) & $C_P$ (J/mol/K) & $\gamma$ \\
0.4 &    7.3168 &    0.6416 &    1.4881 &    1.4997 &  150.7431 &  151.9216 &    1.2185 \\
 0.6 &    8.1998 &    0.4866 &    2.0491 &    2.0609 &  144.5015 &  145.3345 &    1.1846 \\
 0.8 &    8.9406 &    0.3922 &    2.5947 &    2.6065 &  139.1362 &  139.7647 &    1.1516 \\
 1.0 &    9.5895 &    0.3278 &    3.1296 &    3.1411 &  134.3957 &  134.8892 &    1.1204 \\
 1.2 &   10.1728 &    0.2807 &    3.6564 &    3.6676 &  130.1338 &  130.5324 &    1.0912 \\
 1.4 &   10.7065 &    0.2447 &    4.1767 &    4.1876 &  126.2558 &  126.5845 &    1.0642 \\
 1.6 &   11.2010 &    0.2163 &    4.6916 &    4.7022 &  122.6958 &  122.9716 &    1.0393 \\
 1.8 &   11.6636 &    0.1934 &    5.2019 &    5.2121 &  119.4057 &  119.6404 &    1.0167 \\
 2.0 &   12.0994 &    0.1745 &    5.7081 &    5.7181 &  116.3488 &  116.5510 &    0.9961 \\
 2.2 &   12.5126 &    0.1588 &    6.2108 &    6.2205 &  113.4964 &  113.6727 &    0.9777 \\
 2.4 &   12.9062 &    0.1456 &    6.7103 &    6.7197 &  110.8257 &  110.9808 &    0.9614 \\
 2.6 &   13.2827 &    0.1344 &    7.2069 &    7.2161 &  108.3178 &  108.4556 &    0.9473 \\
 2.8 &   13.6440 &    0.1247 &    7.7008 &    7.7098 &  105.9571 &  106.0807 &    0.9352 \\
 3.0 &   13.9919 &    0.1165 &    8.1923 &    8.2011 &  103.7305 &  103.8423 &    0.9253 \\
\hline
\end{tabular}
\label{quantities}
\end{table*}

\begin{table*}
\tiny
\begin{tabular}{cccccccc}
\hline
\multicolumn{8}{l}{2000 K} \\
$P$ (TPa) & $\rho$ (g/cm$^3$) & $\alpha$ ($10^{-5}$/K) & 
$K_T$ (TPa) & $K_S$ (TPa) & $C_V$ (J/mol/K) & $C_P$ (J/mol/K) & $\gamma$ \\
0.4 &    7.2663 &    0.7201 &    1.4653 &    1.4909 &  168.2160 &  171.1596 &  
  1.2150 \\
 0.6 &    8.1561 &    0.5616 &    2.0245 &    2.0513 &  166.3508 &  168.5546 &  
  1.1795 \\

 0.8 &    8.9015 &    0.4646 &    2.5687 &    2.5961 &  164.6315 &  166.3850 &  
  1.1463 \\
 1.0 &    9.5538 &    0.3980 &    3.1025 &    3.1301 &  163.0297 &  164.4779 &  
  1.1159 \\
 1.2 &   10.1399 &    0.3490 &    3.6285 &    3.6561 &  161.5250 &  162.7522 &  
  1.0884 \\
 1.4 &   10.6759 &    0.3113 &    4.1482 &    4.1757 &  160.1022 &  161.1624 &  
  1.0635 \\
 1.6 &   11.1722 &    0.2814 &    4.6627 &    4.6900 &  158.7499 &  159.6800 &  
  1.0412 \\
 1.8 &   11.6364 &    0.2570 &    5.1727 &    5.1998 &  157.4592 &  158.2855 &  
  1.0211 \\
 2.0 &   12.0736 &    0.2367 &    5.6788 &    5.7057 &  156.2230 &  156.9651 &  
  1.0033 \\
 2.2 &   12.4880 &    0.2197 &    6.1814 &    6.2083 &  155.0355 &  155.7082 &  
  0.9874 \\
 2.4 &   12.8826 &    0.2052 &    6.6810 &    6.7077 &  153.8921 &  154.5070 &  

  0.9734 \\
 2.6 &   13.2600 &    0.1928 &    7.1778 &    7.2044 &  152.7888 &  153.3550 &  
  0.9613 \\
 2.8 &   13.6221 &    0.1820 &    7.6721 &    7.6986 &  151.7220 &  152.2470 &  
  0.9508 \\
 3.0 &   13.9707 &    0.1726 &    8.1639 &    8.1905 &  150.6890 &  151.1789 &  
  0.9419 \\
\hline
\end{tabular}
\end{table*}

\begin{table*}
\tiny
\begin{tabular}{cccccccc}
\hline
\multicolumn{8}{l}{3000 K} \\
$P$ (TPa) & $\rho$ (g/cm$^3$) & $\alpha$ ($10^{-5}$/K) & 
$K_T$ (TPa) & $K_S$ (TPa) & $C_V$ (J/mol/K) & $C_P$ (J/mol/K) & $\gamma$ \\
0.4 &    7.2133 &    0.7416 &    1.4428 &    1.4818 &  171.7610 &  176.4066 &    1.2156 \\
 0.6 &    8.1095 &    0.5811 &    2.0001 &    2.0412 &  170.9019 &  174.4187 &    1.1804 \\
 0.8 &    8.8593 &    0.4831 &    2.5428 &    2.5851 &  170.1010 &  172.9292 &    1.1473 \\
 1.0 &    9.5149 &    0.4158 &    3.0755 &    3.1184 &  169.3474 &  171.7070 &    1.1170 \\

 1.2 &   10.1036 &    0.3663 &    3.6006 &    3.6437 &  168.6328 &  170.6524 &    1.0897 \\
 1.4 &   10.6416 &    0.3283 &    4.1195 &    4.1627 &  167.9514 &  169.7131 &    1.0651 \\
 1.6 &   11.1397 &    0.2980 &    4.6335 &    4.6767 &  167.2985 &  168.8586 &    1.0430 \\
 1.8 &   11.6054 &    0.2734 &    5.1431 &    5.1862 &  166.6706 &  168.0694 &    1.0232 \\
 2.0 &   12.0439 &    0.2530 &    5.6489 &    5.6920 &  166.0648 &  167.3322 &    1.0056 \\
 2.2 &   12.4594 &    0.2358 &    6.1515 &    6.1946 &  165.4787 &  166.6375 &    0.9901 \\
 2.4 &   12.8550 &    0.2211 &    6.6511 &    6.6941 &  164.9106 &  165.9786 &    0.9764 \\
 2.6 &   13.2332 &    0.2085 &    7.1479 &    7.1911 &  164.3586 &  165.3502 &    0.9645 \\
 2.8 &   13.5961 &    0.1976 &    7.6424 &    7.6856 &  163.8216 &  164.7481 &    0.9542 \\
 3.0 &   13.9454 &    0.1880 &    8.1345 &    8.1779 &  163.2982 &  164.1692 &    0.9455 \\
\hline
\end{tabular}
\end{table*}

\begin{table*}
\tiny
\begin{tabular}{cccccccc}
\hline
\multicolumn{8}{l}{4000 K} \\
$P$ (TPa) & $\rho$ (g/cm$^3$) & $\alpha$ ($10^{-5}$/K) & 
$K_T$ (TPa) & $K_S$ (TPa) & $C_V$ (J/mol/K) & $C_P$ (J/mol/K) & $\gamma$ \\
0.4 &    7.1595 &    0.7539 &    1.4206 &    1.4728 &  173.0168 &  179.3672 &    1.2171 \\
 0.6 &    8.0621 &    0.5911 &    1.9760 &    2.0312 &  172.5275 &  177.3495 &    1.1820 \\
 0.8 &    8.8162 &    0.4919 &    2.5172 &    2.5741 &  172.0695 &  175.9597 &    1.1490 \\
 1.0 &    9.4750 &    0.4240 &    3.0487 &    3.1065 &  171.6370 &  174.8931 &    1.1187 \\
 1.2 &   10.0662 &    0.3740 &    3.5728 &    3.6311 &  171.2256 &  174.0212 &    1.0913 \\
 1.4 &   10.6063 &    0.3356 &    4.0910 &    4.1496 &  170.8320 &  173.2783 &    1.0667 \\
 1.6 &   11.1062 &    0.3051 &    4.6044 &    4.6631 &  170.4537 &  172.6268 &    1.0446 \\
 1.8 &   11.5732 &    0.2803 &    5.1136 &    5.1723 &  170.0888 &  172.0431 &    1.0248 \\
 2.0 &   12.0130 &    0.2597 &    5.6192 &    5.6779 &  169.7359 &  171.5118 &    1.0073 \\
 2.2 &   12.4296 &    0.2423 &    6.1216 &    6.1804 &  169.3935 &  171.0220 &    0.9918 \\
 2.4 &   12.8261 &    0.2276 &    6.6211 &    6.6800 &  169.0608 &  170.5660 &    0.9781 \\
 2.6 &   13.2052 &    0.2149 &    7.1180 &    7.1771 &  168.7367 &  170.1380 &    0.9662 \\
 2.8 &   13.5688 &    0.2039 &    7.6126 &    7.6720 &  168.4207 &  169.7336 &    0.9559 \\
 3.0 &   13.9187 &    0.1943 &    8.1050 &    8.1647 &  168.1120 &  169.3493 &    0.9472 \\
\hline
\end{tabular}
\end{table*}

\begin{table*}
\tiny
\begin{tabular}{cccccccc}
\hline
\multicolumn{8}{l}{5000 K} \\
$P$ (TPa) & $\rho$ (g/cm$^3$) & $\alpha$ ($10^{-5}$/K) & 
$K_T$ (TPa) & $K_S$ (TPa) & $C_V$ (J/mol/K) & $C_P$ (J/mol/K) & $\gamma$ \\
0.4 &    7.1054 &    0.7635 &    1.3988 &    1.4639 &  173.5974 &  181.6739 &    1.2187 \\
 0.6 &    8.0142 &    0.5983 &    1.9522 &    2.0213 &  173.2826 &  179.4197 &    1.1839 \\
 0.8 &    8.7726 &    0.4979 &    2.4918 &    2.5632 &  172.9873 &  177.9433 &    1.1508 \\
 1.0 &    9.4346 &    0.4292 &    3.0221 &    3.0947 &  172.7079 &  176.8607 &    1.1205 \\
 1.2 &   10.0283 &    0.3788 &    3.5452 &    3.6186 &  172.4417 &  176.0112 &    1.0930 \\
 1.4 &   10.5705 &    0.3400 &    4.0626 &    4.1364 &  172.1866 &  175.3138 &    1.0683 \\
 1.6 &   11.0721 &    0.3092 &    4.5754 &    4.6494 &  171.9410 &  174.7222 &    1.0461 \\
 1.8 &   11.5406 &    0.2842 &    5.0842 &    5.1583 &  171.7039 &  174.2080 &    1.0263 \\
 2.0 &   11.9816 &    0.2634 &    5.5895 &    5.6637 &  171.4742 &  173.7524 &    1.0087 \\
 2.2 &   12.3993 &    0.2459 &    6.0917 &    6.1661 &  171.2511 &  173.3426 &    0.9931 \\
 2.4 &   12.7967 &    0.2311 &    6.5912 &    6.6657 &  171.0341 &  172.9694 &    0.9794 \\
 2.6 &   13.1766 &    0.2183 &    7.0882 &    7.1630 &  170.8225 &  172.6260 &    0.9675 \\
 2.8 &   13.5409 &    0.2072 &    7.5829 &    7.6581 &  170.6158 &  172.3074 &    0.9571 \\
 3.0 &   13.8914 &    0.1975 &    8.0755 &    8.1511 &  170.4138 &  172.0096 &    0.9483 \\
\hline
\end{tabular}
\end{table*}

\begin{table*}
\tiny
\begin{tabular}{cccccccc}
\hline
\multicolumn{8}{l}{6000 K} \\
$P$ (TPa) & $\rho$ (g/cm$^3$) & $\alpha$ ($10^{-5}$/K) & 
$K_T$ (TPa) & $K_S$ (TPa) & $C_V$ (J/mol/K) & $C_P$ (J/mol/K) & $\gamma$ \\
 0.4 &    7.0510 &    0.7720 &    1.3774 &    1.4552 &  173.9111 &  183.7421 &    1.2205 \\
 0.6 &    7.9662 &    0.6044 &    1.9286 &    2.0116 &  173.6920 &  181.1609 &    1.1858 \\
 0.8 &    8.7288 &    0.5027 &    2.4667 &    2.5524 &  173.4862 &  179.5184 &    1.1527 \\
 1.0 &    9.3940 &    0.4333 &    2.9956 &    3.0830 &  173.2912 &  178.3471 &    1.1223 \\
 1.2 &    9.9902 &    0.3823 &    3.5177 &    3.6061 &  173.1051 &  177.4525 &    1.0947 \\
 1.4 &   10.5345 &    0.3432 &    4.0344 &    4.1233 &  172.9267 &  176.7368 &    1.0699 \\
 1.6 &   11.0377 &    0.3122 &    4.5465 &    4.6358 &  172.7549 &  176.1447 &    1.0477 \\
 1.8 &   11.5077 &    0.2869 &    5.0549 &    5.1443 &  172.5888 &  175.6421 &    1.0278 \\
 2.0 &   11.9499 &    0.2659 &    5.5598 &    5.6494 &  172.4278 &  175.2069 &    1.0101 \\
 2.2 &   12.3686 &    0.2483 &    6.0619 &    6.1517 &  172.2713 &  174.8237 &    0.9944 \\
 2.4 &   12.7670 &    0.2333 &    6.5613 &    6.6513 &  172.1190 &  174.4817 &    0.9806 \\
 2.6 &   13.1477 &    0.2204 &    7.0583 &    7.1487 &  171.9703 &  174.1732 &    0.9685 \\
 2.8 &   13.5128 &    0.2092 &    7.5532 &    7.6440 &  171.8252 &  173.8920 &    0.9581 \\
 3.0 &   13.8639 &    0.1995 &    8.0460 &    8.1374 &  171.6831 &  173.6338 &    0.9492 \\
\hline
\end{tabular}
\end{table*}

\begin{table*}
\tiny
\centering
\caption{ Structural parameters of $Pc$-type Mg$_3$SiO$_5$,
$Cm$-type MgSi$_3$O$_5$, and
Gd$_2$S$_3$-type MgSiO$_3$ at 1.5 TPa.}
\label{struct_param2}

\begin{tabular}{ccc}
\multicolumn{3}{c}{$Pc$-type Mg$_3$SiO$_5$} \\
\multicolumn{2}{c}{$(a,b,c)$(\AA)} & (3.9812, 3.9072, 3.8283) \\
\multicolumn{2}{c}{$\beta$ (${}^\circ$)} & 113.02 \\
\hline
Mg$_1$ & ($2a$) & (-0.0010, 0, 0.0030) \\
Mg$_2$ & ($2a$) & (-0.0010, 0, 0.5030) \\
Mg$_3$ & ($2a$) & (0.3999, 0, 0.3238) \\
Mg$_4$ & ($2a$) & (0.3999, 0, 0.8238) \\
Mg$_5$ & ($2a$) & (0.0087, 0, -0.2170) \\
Mg$_6$ & ($2a$) & (0.0087, 0, 0.2830) \\
Si$_1$ & ($2a$) & (-0.4057, 0, -0.0432) \\
Si$_2$ & ($2a$) & (-0.4057, 0, 0.4568) \\
O$_1$  & ($2a$) & (0.3086, 0, -0.3949) \\
O$_2$  & ($2a$) & (0.3086, 0, 0.1051) \\
O$_3$  & ($2a$) & (-0.3075, 0, 0.1891) \\
O$_4$  & ($2a$) & (-0.3075, 0, 0.6891) \\
O$_5$  & ($2a$) & (0.2745, 0, 0.4173) \\
O$_6$  & ($2a$) & (0.2745, 0, 0.9173) \\
O$_7$  & ($2a$) & (-0.2738, 0, -0.1374) \\
O$_8$  & ($2a$) & (-0.2738, 0, 0.3626) \\
O$_9$  & ($2a$) & (-0.0108, 0, -0.3629) \\
O$_{10}$  & ($2a$) & (-.0108, 0, 0.1371) \\
\hline
\end{tabular}
\end{table*}

\begin{table*}
\tiny
\centering
\begin{tabular}{ccc}
\multicolumn{3}{c}{$Cm$-type MgSi$_3$O$_5$ } \\
\multicolumn{2}{c}{$(a,b,c)$(\AA)} & (8.6355, 1.9901, 7.7799) \\
\multicolumn{2}{c}{$\beta$ (${}^\circ$)} & 90.40 \\
\hline
Mg$_1$ & ($2a$) & (0.9998, 0, 0.9991) \\
Mg$_2$ & ($2a$) & (0.5150, 0, 0.4708) \\
Si$_1$ & ($2a$) & (0.8660, 0, 0.7618) \\
Si$_2$ & ($2a$) & (0.2776, 0, 0.4971) \\
Si$_3$ & ($2a$) & (0.2443, 0, 0.9767) \\
Si$_4$ & ($2a$) & (0.6523, 0, 0.2408) \\
Si$_5$ & ($2a$) & (0.6104, 0, 0.7176) \\
Si$_6$ & ($2a$) & (0.9019, 0, 0.2448) \\
O$_1$  & ($2a$) & (0.8877, 0, 0.4373) \\
O$_2$  & ($2a$) & (0.0241, 0, 0.8014) \\
O$_3$  & ($2a$) & (0.0521, 0, 0.3239) \\
O$_4$  & ($2a$) & (0.1450, 0, 0.1155) \\
O$_5$  & ($2a$) & (0.6835, 0, 0.4200) \\
O$_6$  & ($2a$) & (0.2330, 0, 0.7895) \\
O$_7$  & ($2a$) & (0.0967, 0, 0.5824) \\
O$_8$  & ($2a$) & (0.8197, 0, 0.0818) \\

O$_9$  & ($2a$) & (0.4828, 0, 0.1479) \\
O$_{10}$ & ($2a$) & (0.6291, 0, 0.9339) \\
O$_{11}$ & ($2a$) & (0.4257, 0, 0.6394) \\
O$_{12}$ & ($2a$) & (0.2858, 0, 0.2778) \\
O$_{13}$ & ($2a$) & (0.3955, 0, 0.9056) \\
O$_{14}$ & ($2a$) & (0.7589, 0, 0.6228) \\
\hline
\end{tabular}

\begin{tabular}{ccc}
\multicolumn{3}{c}{Gd$_2$S$_3$-type ($Pbnm$) MgSiO$_3$} \\
\multicolumn{2}{c}{$(a,b,c)$(\AA)} & (5.5174, 2.0286, 5.4829) \\
\hline
Mg & ($4c$) & (0.2997, 0.2500, 0.0039) \\
Si & ($4c$) & (0.4836, 0.2500, -0.3046) \\
O$_1$ & ($4c$) & (-0.2718, 0.2500, -0.2087) \\
O$_2$ & ($4c$) & (0.0473, 0.2500, 0.3734) \\
O$_3$ & ($4c$) & (0.3905, 0.2500, 0.4377) \\
\hline
\end{tabular}

\end{table*}

\end{document}